# The Pioneer Anomaly

**José A. de Diego and Darío Núñez**

*Instituto de Astronomía, Universidad Nacional Autónoma de México*
*Ciudad Universitaria, 04510 México, D.F., México*
jdo@astroscu.unam.mx

## ABSTRACT

Analysis of the radio-metric data from Pioneer 10 and 11 spacecrafts has indicated the presence of an unmodeled acceleration starting at 20 AU, which has become known as the Pioneer anomaly. The nature of this acceleration is uncertain. In this paper we give a description of the effect and review some relevant mechanisms proposed to explain the observed anomaly. We also discuss on some future projects to investigate this phenomenon.

## RESUMEN

El análisis de datos radiométricos de las naves espaciales Pioneer 10 y 11 ha indicado la presencia de una aceleración inexplicada a partir de 20 UA, que se ha dado en denominar la anomalía del Pioneer. La naturaleza de esta aceleración es incierta. En este artículo damos una descripción del efecto y hacemos una reseña de los mecanismos más relevantes que se han propuesto para explicar la anomalía observada. También discutimos algunos proyectos a futuro para investigar este fenómeno.

## INTRODUCTION

Radiometric data from the Pioneer 10 and 11, Galileo, Ulysses and Cassini spacecrafts have brought out the existence of an anomalous acceleration towards the Sun with a value of $a_A = (8.74\pm1.33)\times10^{-10}$ m/s$^2$ (Anderson et al., 2002a; 2003). The phenomenon has resisted any attempt for explanation invoking either failures in the tracking algorithm, engineering causes or external forces acting upon the space probes.

Both Pioneer spacecrafts follow approximate opposite escape hyperbolic trajectories near the plane of the ecliptic on opposite sides of the Solar System. Galileo crashed into Jupiter on September 21, 2003. Ulysses has flown over the Sun's poles for the third time in 2007 and 2008; as its aging radioisotope generators continue to run down the mission is coming to an end after 18 years. Cassini is still orbiting around Saturn and the mission will be extended probably for another two years. Thus, the five spacecrafts had quite different trajectories as well as four unlike physical designs. These facts suggest that the cause of the anomaly may be found outside the engineering design of the spacecrafts, motivating the search of explanations in external forces and new physical laws.

Pioneer 10 was launched from Cape Canaveral Air Force Station on March 2, 1972 and Pioneer 11 was launched on April 5, 1973. Pioneer 10 was the first spacecraft to travel through the asteroid belt and to make direct observations of the planet Jupiter, while Pioneer *11* was the first spacecraft to explore the planet Saturn. Pioneer 10 is currently in the outer edges of the solar system, within the Sun's heliopause. The last communications with Pioneer *11* occurred in November 1995. The last signals from Pioneer *10* occurred in January 2003. As of February 13, 2008, Pioneer 10 was about 95.07 AU from the Sun (see Figure 1).

Pioneer 10 science mission ended on March 31, 1997. Since then, the weak signal has been tracked by the NASA's Deep Space Network for communication technology studies and supporting future interstellar probe missions. Pioneer 10 eventually became the first man-made object to leave the solar system. During the mission, the power source degraded, thus the last telemetry data was obtained on April 27, 2002 (80 AU from the Sun), and the last three contacts had very faint signal and the telemetry was not received; finally the signal was no longer detected, and the last contact attempt was made on 7 February 2003.

Pioneer 11 sent the last coherent Doppler data on October 1, 1990 (30 AU from the Sun). In September 1995 (6.5 billion km from Earth), Pioneer 11 could no longer make any scientific observations. On September 30, 1995 the routine daily mission operations were stopped. Intermittent contact continued until November 1995. The antenna is misaligned and the spacecraft cannot be maneuvered to point back at the Earth. As of February 13, 2008, Pioneer 11 was about 75.53 AU from the Sun (see Figure 1).

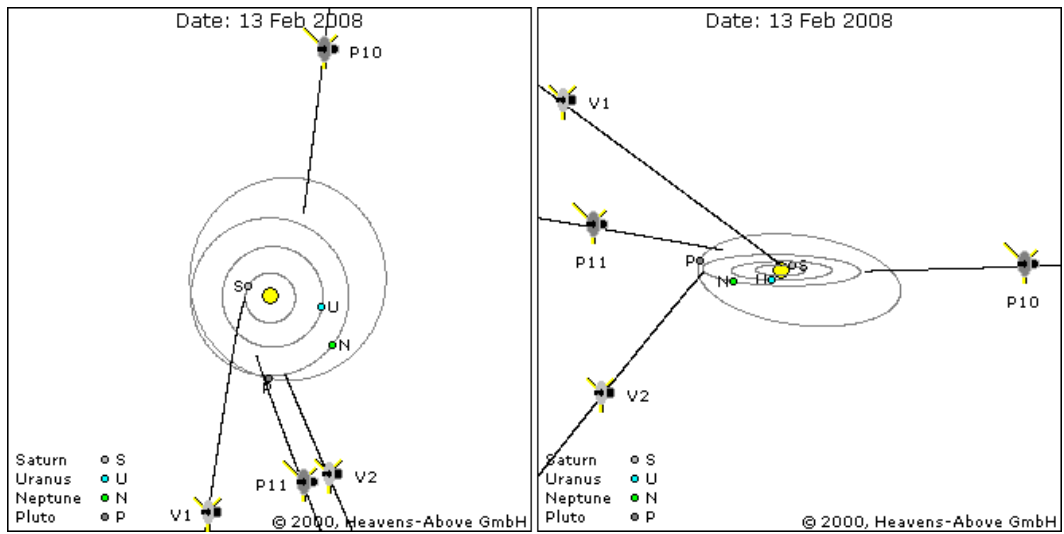

**Figure 1. Pioneers 10 and 11, and Voyagers 1 and 2 positions and trajectories up to February 13, 2008 (Heavens Above: http://www.heavens-above.com).**

The Pioneers have the most precise navigation system for any deep space vehicle up to date (Null 1976). They had been designed for conducting precision celestial mechanics experiments. For these purposes, the spacecrafts count with a mHz precision Doppler tracking (they had an acceleration sensitivity of ~$10^{-10}$ m/s$^2$), an advanced attitude control (spin-stabilized), and Radioisotope Thermoelectric Generators (RTGs) attached on extended arms to help to the stability and to minify thermal processes which may affect the crafts.

## THE ANOMALOUS ACCELERATION

The spacecraft trajectories have been modeled from radiometric data, considering conventional forces acting upon the spacecrafts, either gravitational or non-gravitational. Figure 2 shows that starting at 20 AU from the Sun, the accuracies of the models were hampered by a small Doppler frequency blue-shifted drift (Anderson et al. 1998, 2002a; however, the onset of the anomaly may occur around 10 AU, near the orbit of Saturn). This drift [(5.99±0.01)×$10^{-9}$ Hz/s] has been interpreted by Anderson et al. (2002a) as a constant, anomalous acceleration towards the Sun of (8.74±1.33)×$10^{-10}$ m/s$^2$, but its origin is unknown. Nonetheless, Anderson et al. (2002a) also consider a constant time deceleration of (2.92±0.44)×$10^{-18}$ s/s$^2$ as a possible origin of the anomaly.

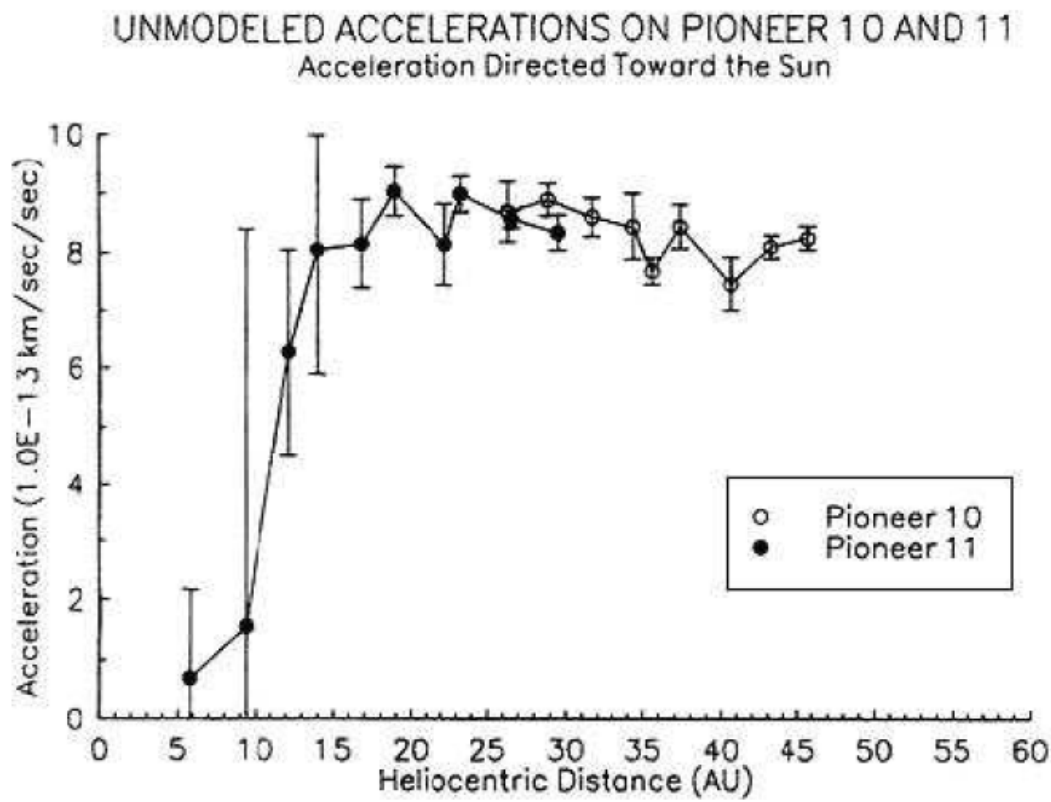

Figure 2. Unmodeled accelerations on Pioneer 10 and 11. The acceleration starts near Uranus, around 20 AU, but the onset of the perturbation may have started near Saturn, around 10 AU. Figure extracted from Anderson et al. 2002a.

In their excellent review, Turyshev et al. (2006) itemize the following independent studies with four different navigational softwares that show the presence of the anomaly for both Pioneer 10 and 11:

- The JPL's Orbit Determination Program, developed in 1980-2005
- The Aerospace Corporation's CHASMP code (Anderson et al. 1998; 2002a)
- The Goddard Space Flight Center code (Markwardt 2002)
- The code developed at the Institute of Theoretical Astrophysics, University of Oslo, Norway (Olsen 2007)

The origin of the anomaly has been attributed to both, intrinsic and extrinsic causes. In the next two sections we will review the possible effects that may accelerate the spacecrafts.

## ON-BOARD CAUSES

Some on-board effects have been proposed to explain the anomaly (Anderson et al. 2002a-b). These effects include electromagnetic emission from the spacecrafts (either thermal losses or beamed from the spacecraft antennae) and gas leaks. Figure 3 shows the main external features of the Pioneers.

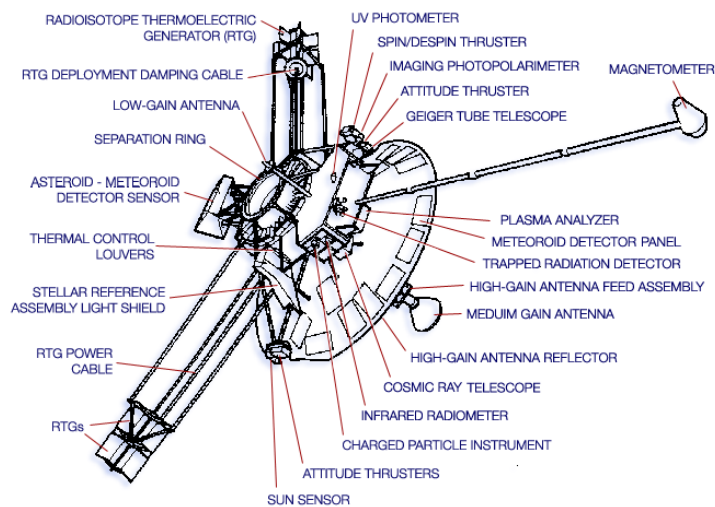

Figure 3. Pioneers 10 and 11 spacecrafts (NASA, http://www2.jpl.nasa.gov/basics/pioneer.html).

The effect that may cause the largest bias is the reaction force on the spacecraft antenna exerted by the radio beam, which may account for an acceleration of $(1.1\pm0.1) \times 10^{-10}$ m/s$^2$. However, as the radio beam points to the Earth, the force exerted by the radio beam is in the opposite direction to the observed acceleration, and thus the anomaly would be even larger.

Another possibility is that the heat emitted from the RTGs is anisotropically reflected by the spacecraft high gain antennae (Anderson et al. 1998, 2002a). At the beginning of the missions, the heat sources produced almost 2.5 kW of heat, while to account for the observed anomaly, only about 64 W of directed constant heat is needed. However, Anderson et al. (2002a) have calculated that only 4 W of directed power could be generated, based on the designs of the spacecrafts and the RTGs. These authors estimate that the heat reflection may account for $(-0.55 \pm 0.55) \times 10^{-10}$ m/s$^2$ of the observed anomaly. Furthermore, the radiant heat from the RTGs decreases by the decay of the radioactive fuel, and thus the amount of the acceleration should decrease with time. It is worth noting that models to explain the thermal emission from the spacecrafts are complicated and subject to large uncertainties, hence the thermal effect may reach a value of 24 W and account for $-3.3 \times 10^{-10}$ m/s$^2$ (Scheffer 2003). However, Olsen (2007) investigates the temporal variations of the anomalous acceleration and concludes that the Doppler data for Pioneer 10 and 11 cannot be used to distinguish between a constant acceleration and an acceleration proportional to the remaining plutonium in the RTGs.

The inner sides of the RTGs received the solar wind during the early parts of the missions, while the outer sides received the impact of the solar system dust particles. Both effects can degrade the surfaces of the RTGs and produce asymmetries in the heat radiated away from the RTGs in the fore and aft directions. Anderson et al. (2002a) have estimated an upper limit to the uncertainty of the asymmetric emissivity of the RTGs of 6.12 W, which implies a contribution to the uncertainty of the anomalous acceleration of $0.85 \times 10^{-10}$ m/s$^2$. However, as these authors point out, this mechanism also depends on the radioactive decay and thus a decrease of the acceleration would be expected, but it has not been observed.

The heat produced by the electrical power has also been investigated as a possible mechanism to produce the anomalous acceleration (Murphy 1999), which may be redirected through the closed thermal control louvers (Scheffer 2003). However, the electrical power generated in the spacecrafts also depends on the radioactive decay, and the resulting acceleration would decrease in time.

A contribution to the anomalous acceleration due to gas expulsion from the spacecrafts has also been considered. There are two possible sources for this gas loss, the Helium from the RTGs, and the gas from the propulsion system. The He in the RTGs is the result of the α-decay of $^{238}$Pu. To account for the anomaly, the He gas should leave the RTGs in one direction at a rate of 0.77 g/yr. But Anderson et al (2002a) have ruled out this mechanism and give a value for the estimated acceleration contribution of $(0.15 \pm 0.16) \times 10^{-10}$ m/s$^2$. On the other hand, Anderson et al (2202a) have also estimated the acceleration uncertainty due to gas leakage from the propulsion system ($\pm 0.56 \times 10^{-10}$ m/s$^2$), and they have concluded that this mechanism is very unlikely to explain the anomaly because it is an stochastic process that would affect Pioneers 10 and 11 differently. Table 1 summarizes the intrinsic effects proposed for the Pioneer anomaly

Table 1. Summary of the intrinsic effects proposed for the Pioneer anomaly

| Responsible | Effects | $10^{-10}$ m/s$^2$ | Commentary |
|---|---|---|---|
| Antenna | Radio signal | 1.10±0.10 | Opposite sign |
| Radioisotope Thermoelectric Generators | Antenna reflection | -0.55±0.55 | Decreases |
| Radioisotope Thermoelectric Generators | Anisotropic emission | ±0.85 | |
| Radioisotope Thermoelectric Generators | He Expulsion | 0.15±0.16 | $\alpha$-decay $^{238}$Pu |
| Propulsion system | Gas leakage | ±0.56 | |
| Electrical circuits | Electrical heat | | Non constant |

## EXTRINSIC CAUSES

Both Pioneers have basically the same physical features, but the Galileo spacecraft that explored the Jovian system, the Ulysses heliosphere observatory and the Cassini probe orbiting Saturn have quite different designs. Besides, the five spacecraft had very different trajectories, and even so all of them showed evidence for the same anomalous acceleration. Thus, it is difficult to explain why four different designs and five different trajectories would all endure different effects such as heat radiation, antennae radio emission, and gas leaks, but still yielding the same result.

This difficulty to find an onboard effect to account for the anomalous acceleration, and that the same acceleration is suggested for all the spacecrafts despite their trajectories and differences in their design, has boosted a variety of imaginative hypotheses, ranging from evidence of external forces acting on the spacecrafts, to *new physics*. Although we will focus on some conventional external forces invoked to accelerate the spacecrafts, we will also present an outline of the new physics hypothesis.

### New Physics

Anderson et al. (1998) considered an *unknown* interaction of the radio signals with the solar wind. Foot & Volkas (2001) consider *mirror* matter, a kind of dark matter that would restore the parity symmetry. Modified-inertia approaches have been considered under the Modified Newtonian Dynamics theory by Milgrom (2001) and under Unruh radiation by McCulloch (2006). Cadoni (2004) studied the coupling of gravity with a scalar field with an exponential potential, while Bertolami & Páramos (2004) also applied a scalar field in the framework of braneworld scenarios. Jaekel & Reynaud (2005) discuss another solution in terms of the parametrized post-Newtonian formalism. Gravitational coupling resulting in an increase of the constant G with scale is analyzed by Bertolami & García-Bellido (1996). Moffat (2004) proposes an explanation in terms of a nonsymmetric gravitational theory. Rañada (2005) discusses the effect of a background gravitational potential that pervades the universe and is increasing because of the expansion, provoking a drift of clocks (see also Anderson et al. 2002a); however, such an effect has not been observed in radio signals from pulsars (Matsakis et al. 1997; Wex 2001). For Ostvang (2002), cosmic expansion applies directly to gravitationally bound systems according to the so-called *quasi-metric* framework. The scale factor of the space-time background would cause an anomaly in the frequency according to Rosales (2004; 2005). The cosmological constant has also been invoked to produce acceleration by Nottale (2008) and a gravitational frequency shift by Mbelek (2004)

### Conventional forces

There are a number of effects due to conventional forces that have been proposed by different authors. Thus, Nieto et al. (2005) investigated the drag force due to interplanetary dust and calculated that the density of dust necessary to provoke the acceleration would be five orders of magnitude larger than the density estimated for the Kuiper belt dust (~$10^{-24}$ g/cm$^3$); a similar result was obtained by Bertolami & Vieira (2006). Bini et al. (2004) considered a nongravitational acceleration of the sun, orthogonal to the ecliptic, but they found that it is necessary that the sun would emit all the electromagnetic radiation in the opposite direction. Lämmerzahl et al. (2006) have studied the coincidence of the Pioneer anomalous acceleration with the value *cH*, where *c* is the velocity of light and *H* the Hubble constant, and the possible influences on the signal propagation, trajectory of the spacecraft, magnitude of the gravitational field and the definition of the astronomical unit; however, these authors calculate a value of only *vH*, i.e. a factor *v/c* less than the observed anomalous acceleration, where *v* is the spacecraft velocity.

Anderson et al. (2002a) considered a gravitational source in the Solar System as a possible source for the anomaly. According to the equivalence principle, such a gravitational source would also affect the motions of the planets. In the case of the inner planets, that have orbits determined with great accuracy, they show no evidence for the expected anomalous motion if the source of the anomaly were located in the inner Solar System. For example, in the case of Mars, range data provided by the Mars Global Surveyor and Mars Odyssey missions have yielded measurements of the Mars system center-of-mass relative to the Earth to an accuracy of one meter (Konopliv et al. 2006). However, the anomaly has been detected beyond 20 AU (i.e., beyond Uranus, 19 AU), and the orbits of the outer planets have been determined only by optical methods, resulting in much less accurate planet ephemerides. Thus, Page et al. (2006) conclude that such anomalous gravitational disturbance would not be detected in the orbits of the outer planets. However, the same authors (Wallin et al. 2007) test the orbits of 24 Trans-Neptunian Objects using bootstrapping analysis and find no evidence of the anomaly.

Other authors have also argued that a gravitational effect would have observable evidence on the orbits of the outer planets, which is not observed. Hence, Izzo and Rathke (2005) used parametric constraints to the orbits of Uranus and Neptune and found that the reduced Solar mass would not be compatible with the measurements. A similar result was obtained by Iorio and Giudice (2006) based on the Gauss equations to estimate the effect of a gravitational perturbation in terms of the time rate of change on the osculating orbital elements. For these authors, the perturbation would produce long-period, secular rates on the perihelion and the mean anomaly, and short-period effects on the semimajor axis, the eccentricity, the perihelion and the mean anomaly large enough to be detected. Tangen (2007) also considers the effect on the path of the outer planets by a disturbance on a spherically symmetric space-time metric, and rules out any model of the anomaly that implies that the Pioneer spacecrafts move geodesically in a perturbed space-time metric.

Despite these suggesting pieces of work, the lack of evidence for perturbations in the orbits of the outer planets is not conclusive, as they are based on statistical analysis relying on some simplified models, rather than on direct measurements as in the case of the inner planets. Thus, the possibility of a gravitational perturbation on the Pioneer paths remains still possible and has been considered by Nieto (2005) and Bertolami & Vieira (2006), who studied the possible gravitational effects produced by different Kuiper-Belt mass distributions, concluding that the Kuiper-Belt cannot produce the observed acceleration. On the other hand, de Diego et al. (2006) also discarded the gravitational attraction by the Kuiper-Belt, but they proposed that the observed deceleration in the Voyager space probes could be simply explained by the gravitational pull of a distribution of matter, and they pointed out that, as this matter has not been detected yet, it could be the Solar system dark matter. Thus, considering a Navarro, Frenk & White (1997) dark matter distribution, de Diego et al. (2006) show that there should be several hundreds earth masses of dark matter in the Solar System. For these authors, a spatial distribution of part of this dark matter as the dust contained in the Kuiper-Belt could explain the observed anomaly. The dark matter gravitational pull has been recently considered by Nieto (2008) who proposes the analysis of the New Horizons spacecraft data when the probe crosses the orbit of Saturn (see next section).

## NEXT STEPS

Since the discovery of the Pioneer anomaly, there have been interesting proposals to launch a mission to investigate this phenomenon. For example, Dittus et al (2005) propose a dedicated mission based on a *formation-flying approach* that relies on an actively controlled spacecraft and a set of passive test-masses. On the other hand, Rathke & Izzo (2005) propose a non-dedicated mission, either a planetary exploration spacecraft or a piggybacked micro-spacecraft to be launched from a mother spacecraft travelling to Saturn or Jupiter. Several technological goals have been planned for such missions, including positioning control, thermal design, control of the antennae emission, etc.

However, to launch a mission dedicated *per se* to test the anomaly could be quite embarrassing, as long as there is still the possibility that the internal effects were the responsible for the observed behavior. Nonetheless, the possibility of an external influence is so intriguing that any future Solar System mission should have the capabilities to test the anomaly, and it is also worth studying the possibility to test the anomaly using current Solar System probes. In this sense, a recent proposal by Nieto (2008) consists on analyzing the data from the New Horizons spacecraft traveling to Pluto and the Kuiper-Belt. This spacecraft was launched on 19 January 2006, and on its pass through the orbit of Saturn in mid-2008 could supply a clear test of an onset of a Pioneer-like anomaly, as suggested by the Pioneer data. In the future, by increasing the accuracy on the position and velocities of the comets (probably landing probes with telemetric capabilities on their surfaces), they could also be used to test the external effects on their motion within large regions of the Solar System.

Previous to any realistic proposal to launch a dedicated mission to study the Pioneer Anomaly, it is necessary to study the complete Pioneer database in order to rule out, as far as possible, any intrinsic cause, as well as to obtain clever insights on the phenomenon. In this sense, a remarkable effort to rescue and analyze early Pioneer data (before 1987) is currently in progress (Toth & Turyshev 2008). These data include telemetric measurements as well as the state of the spacecraft instruments (temperatures, currents and voltages, gas pressure). A careful analysis of the thermal and gas losses from the spacecrafts will help in modeling intrinsic possible causes for the anomaly, while the telemetric data possibly will bring new light on the onset of the anomaly. These data will also be very useful to discriminate the direction of the perturbation, and thus the possible origin of the anomaly. Thus, if the anomaly is directed to the Sun it would suggest a Solar or Solar System origin; if directed to the Earth it would be probably associated with the frequency standards; if the anomaly is in the direction of the spacecraft motion, it would be related to inertial o drag forces; and if the anomaly is linked with the direction of the rotational axis, it would be a strong evidence for intrinsic spacecraft causes.

## CONCLUSIONS

In this paper we have reviewed the Pioneer Anomaly, an unmodeled acceleration of $a_A = (8.74\pm1.33)\times10^{-10}$ m/s$^2$ towards the sun detected in radiometric data from the Pioneer 10 and 11 spacecrafts, and also suggested in the radiometric data from Galileo, Ulysses and Cassini spacecrafts. The origin of the anomaly is still uncertain, but different explanations have been proposed. Errors in the navigational software used to calculate the trajectories of the Pioneers have been ruled out after four different studies have shown evidence of the same anomalous effect. On-board causes such as heat radiation or gas leaks cannot be completely ruled out, but it is difficult to explain the presence of the same effect in five spacecrafts with different designs and trajectories. Hence, extrinsic causes for the origin of the anomaly have been studied. Gravitational disturbances and other conventional forces acting upon the spacecrafts have been invoked by different researchers. Although the presence of dark matter in the outer Solar System may be a firm candidate for the origin of the anomalous acceleration, the difficulty to test its effect on the orbits of the planets beyond Saturn, and the lack of evidence of disturbances in the ephemeris data for Uranus, Neptune and Pluto, prevent to reach a final conclusion about any gravitational effect.

Several groups have hypothesized that the observed anomaly might be a consequence of the incompleteness of the current theory of gravitation, or a signature of new physical phenomena. Although these hypotheses are highly speculative, they are also very exciting because of the importance that they can have in our understanding of the physical laws if any of them is confirmed. The human curiosity impels the scientific community to continue the quest for an explanation to the Pioneer Anomaly. Hence, new efforts are continuously arising to disentangle this mysterious effect, and if we reach to a limit in our capacity to test the proposed explanations, a future space mission would be necessary to collect the accurate data that we will need to discard or confirm the theories.

## ACKNOWLEDGEMENTS

This work has been supported by the CONACyT grant 50296. Current positions for the Pioneer 10 and 11, and Voyager 1 and 2, have been obtained from *Heavens Above* (http://www.heavens-above.com). This research has made use of NASA's Astrophysics Data System Bibliographic Services.